\documentclass[aps,prx,twocolumn,superscriptaddress]{revtex4}

\usepackage{graphicx}
\usepackage{dcolumn}
\usepackage{bm}
\usepackage{amssymb}
\usepackage{amsfonts}
\usepackage{latexsym}
\usepackage{enumerate}
\usepackage{amsmath}
\usepackage{times}
\usepackage{algorithm}
\usepackage{algorithmic}
\usepackage{cases}
\usepackage{mathrsfs}
\usepackage{color}

\newcommand{\Tr}{\operatorname{Tr}}

\newtheorem{theorem}{Theorem}
\newtheorem{conjecture}{Conjecture}
\newtheorem{proposition}{Proposition}
\newtheorem{lemma}{Lemma}

\newtheorem{corollary}{Corollary}
\newtheorem{definition}{Definition}

\begin{document}
\widetext
\title{Verifying commuting quantum computations via fidelity estimation of weighted graph states}
\author{Masahito Hayashi}
\email{masahito@math.nagoya-u.ac.jp}
\affiliation{Graduate School of Mathematics, Nagoya University, Nagoya 464-8602, Japan}
\affiliation{Shenzhen Institute for Quantum Science and Engineering, Southern University of Science and Technology, Shenzhen 518055, China}
\affiliation{Centre for Quantum Technologies, National University of Singapore, 3 Science Drive 2 117542, Singapore}
\author{Yuki Takeuchi}
\email{takeuchi.yuki@lab.ntt.co.jp}
\affiliation{NTT Communication Science Laboratories, NTT Corporation, 3-1 Morinosato Wakamiya, Atsugi, Kanagawa 243-0198, Japan}

\begin{abstract}
The instantaneous quantum polynomial time model (or the IQP model) is one of promising models to demonstrate a quantum computational advantage over classical computers. If the IQP model can be efficiently simulated by a classical computer, an unlikely consequence in computer science can be obtained (under some unproven conjectures). In order to experimentally demonstrate the advantage using medium or large-scale IQP circuits, it is inevitable to efficiently verify whether the constructed IQP circuits faithfully work. There exists two types of IQP models, each of which is the sampling on hypergraph states or weighted graph states. For the first-type IQP model, polynomial-time verification protocols have already been proposed. In this paper, we propose verification protocols for the second-type IQP model. To this end, we propose polynomial-time fidelity estimation protocols of weighted graph states for each of the following four situations where a verifier can (i) choose any measurement basis and perform adaptive measurements, (ii) only choose restricted measurement bases and perform adaptive measurements, (iii) choose any measurement basis and only perform non-adaptive measurements, and (iv) only choose restricted measurement bases and only perform non-adaptive measurements. In all of our verification protocols, the verifier's quantum operations are only single-qubit measurements. Since we assume no i.i.d. property on quantum states, our protocols work in any situation.
\end{abstract}
\maketitle

\section{Introduction}
Quantum computing is believed to be able to perform several computational tasks faster than classical computing. Indeed, some efficient quantum algorithms that outperform the best known classical algorithms have been found for the integer factorization~\cite{S97}, approximations of Jones polynomials~\cite{AJL09,AAEL07}, and simulations of quantum many-body dynamics~\cite{GAN14}. In addition, quantum computational advantages have been shown in terms of the query complexity~\cite{S94,G97} and the communication complexity~\cite{BCW98,R99}.

Recently, the quantum computational advantage has also been shown in terms of sampling problems, which is called the quantum (computational) supremacy~\cite{HM17}. If an appropriately designed quantum computing model can be efficiently simulated by a classical computer, an unlikely consequence in computer science can be obtained under some unproven conjectures (for details, see Sec.~\ref{VIIIB}). So far, to demonstrate the quantum supremacy, several quantum computing models have been proposed~\cite{TD04,BJS11,TT16,BMS16,AA13,FKMNTT18,M17,BFNV18,TYT14,BFK18,MTN18}. As an advantage of this approach, the quantum computing model do not have to be universal one. Because of this advantage, this approach is considered to be well suited to demonstrate the quantum computational advantage using near-term quantum technologies. Several proof-of-principal small-scale experiments have already been performed towards the demonstration of the quantum supremacy~\cite{BFKDARW13,TDHNSW13,BSVFVLMBGCROS15,WHLSLHDCLQLHSKPHLP17,ZLLPSHHDZLZWYWJLCLLP18,LBAW08}.

\begin{figure}[t]
\includegraphics[width=8cm, clip]{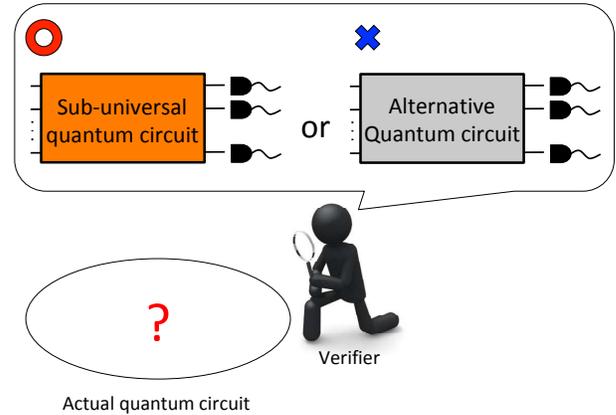}
\caption{Illustration of the verification for the sub-universal model. Given an (experimentally realized) actual quantum circuit, a verifier checks whether the circuit is the target sub-universal circuit (the correctly working device) or an alternative circuit that generates a completely different output probability distribution.}
\label{verification}
\end{figure}

In order to extend these experimental demonstrations of the quantum supremacy to medium or large-scale ones, efficient methods of verifying whether the target sub-universal model is faithfully realized are inevitable 
(see Fig.~\ref{verification}). From this importance, several efficient verification protocols have been proposed for various sub-universal quantum computing models~\cite{HKSE,MSM17,TM18,ZH,FKD18}. However, there is a possibility that conjectures making classical simulations of these verifiable sub-universal models intractable will be rejected. Therefore, it is theoretically and experimentally important to investigate the verifiability of other sub-universal models.

In this paper, we focus on the instantaneous quantum polynomial time (IQP) model~\cite{SB09}. Simply speaking, this model can be considered as a non-adaptive measurement-based quantum computation (MBQC)~\cite{RB01,RBB03}. In other words, in the IQP model, an entangled resource state is prepared, and then each of all qubits is simultaneously measured (for details, see Sec.~\ref{VIIIB}). By appropriately designing the resource state, the IQP model can generate the output probability distribution whose simulation seems to be hard for any classical sampler.
More precisely, if the IQP model can be efficiently simulated by a classical computer, the polynomial-time hierarchy would collapse to its third level, which is an unlikely consequence in computer science, under some unproven conjectures.
In Ref.~\cite{BMS16}, two types of IQP circuits have been proposed, and their hardness of classical simulations have also been shown under different conjectures. The first one is based on hypergraph states~\cite{RHBM13}, which is generalizations of graph states. For this type of IQP circuits, verification protocols have already been proposed via the efficient fidelity estimation of hypergraph states~\cite{HKSE,MSM17,TM18,ZH}. On the other hand, the second type is based on weighted graph states, which are another generalizations of graph states (for the definition, see Sec.~\ref{II}). It was open whether this type of IQP circuits are efficiently verifiable.

In this paper, we affirmatively solve this open problem. More precisely, we propose efficient (polynomial-time) fidelity estimation protocols of weighted graph states for each of the following four situations where a verifier can (i) choose any measurement basis and perform adaptive measurements, (ii) only choose restricted measurement bases and perform adaptive measurements, (iii) choose any measurement basis and only perform non-adaptive measurements, and (iv) only choose restricted measurement bases and only perform non-adaptive measurements. In all of our verification protocols, the verifier's quantum operations are only single-qubit measurements. Applying these protocols, we show that the weighted-graph-state-based IQP model is also verifiable. In other words, we show that the similar unlikely consequence to that of the IQP model is obtained using quantum states that pass our verification protocols. Our fidelity estimation protocols do not assume any independent and identically distributed (i.i.d.) property on quantum states. Therefore, our verification protocols for the IQP model work in any situation. Even when the IQP circuit is given by a malicious server, our protocols correctly verify whether the IQP circuit faithfully works. Furthermore, since the difference between the universal MBQC and the IQP model is only adaptive measurements, our fidelity estimation protocols can also be used for the verification of the MBQC.

The rest of this paper is organized as follows: In Sec.~\ref{II}, as preliminaries, we review the definition of weighted graph states and explain some terminologies that are necessary to understand our result. In Sec.~\ref{S1-2}, we review some known mathematical facts that are used in proofs of our theorems. In Secs.~\ref{S2}, \ref{S3}, \ref{S4}, \ref{VII}, as the main result, we propose four kinds of verification protocols for weighted graph states. In Sec.~\ref{VIII}, we apply our verification protocols to verify the MBQC and the IQP model. Section~\ref{IX} is devoted to the conclusion and discussion.

\section{Weighted graph states}
\label{II}
In this section, we review the definition of weighted graph states~\cite{HDERvB06,HCDB07}.
\begin{definition}[Weighted graph states]
Let $G\equiv(V,E,\Theta)$ be a weighted graph, i.e., a triple of a set $V$ of vertices, a set $E$ of edges, and a set $\Theta\equiv\{\theta_{jk}\}_{j,k=1}^n$ $(j<k)$ of weights, where $n\equiv|V|$. Here, $|V|$ represents the number of vertices, and $\theta_{jk}\in\mathbb{R}$ represents the weight of the edge $(j,k)$. Note that if $(j,k)\notin E$, $\theta_{jk}=0$.
A weighted graph state $|G\rangle$ corresponding to $G$ is defined as
\begin{eqnarray}
|G\rangle\equiv \left[\prod_{(j,k)\in E}\Lambda_{jk}(\theta_{jk})\right]|+\rangle^{\otimes n},
\end{eqnarray}
where each $|+\rangle(\equiv\frac{|0\rangle+|1\rangle}{\sqrt{2}})$ state is placed on each vertex, and
\begin{align*}
\Lambda_{jk}(\theta_{jk})
&\equiv
 |0\rangle\langle 0|_j\otimes I_k+|1\rangle\langle 1|_j\otimes (|0\rangle\langle 0|_k+e^{i\theta_{jk}}|1\rangle\langle 1|_k) \\
&=
 |0\rangle\langle 0|_k\otimes I_j
 +|1\rangle\langle 1|_k\otimes (|0\rangle\langle 0|_j+e^{i\theta_{jk}}|1\rangle\langle 1|_j)
\end{align*}
is the controlled-$Z$ rotation gate acting on the $j$-th and $k$-th qubits. Here, $I_{k(j)}$ is the two-dimensional identity operator on the $k$ $(j)$-th qubit.
\end{definition}

A subset of $ V$ is called an independent set if no two vertices are connected to each other.
A set  $\mathscr{A}=\{A_1, A_2, \ldots, A_m \}$ of independent sets of $V$ is called an independence cover  
if $\cup_{l=1}^m A_l=V$. The cover $\mathscr{A}$ also defines a coloring of $G$ with $m$ colors 
when $\mathscr{A}$ forms a partition of $V$, that is,
when $A_l$ are pairwise disjoint (assuming no $A_l$ is empty).
Hereafter, we consider the independence cover whose entries are pairwise disjoint.
A weighted graph $G$ is $m$-colorable if its vertices can be colored using $m$ different colors such that any two adjacent vertices are assigned with different colors. The chromatic number $\chi(G)$ of $G$ is the minimal number of colors in any coloring of $G$ or, equivalently, the minimal number of elements in any independence cover of $G$. 
In particular, a two-colorable graph is also called a bipartite graph.

\section{Fundamental facts}\label{S1-2}
First, we review fundamental facts for a conventional testing protocol 
based on a non-negative operator $\Omega$ satisfying $I\ge \Omega$ on the single copy system
as follows.

\begin{definition}
The verifier randomly chooses $N$ copies from $N+1$ copies
and apply the same POVM $
\{\Omega,I-\Omega\}$ to each of the $N$ copies.
Then, if all outcomes correspond to $\Omega$, the verifier accepts the remaining single copy $\sigma$.
Otherwise, the verifier rejects it. 
This test is called the $N$-random sampling test of $\Omega$.
When we employ the $N$-random sampling test,
the operator $\Omega$ is called the test operator.
\end{definition}
We here note that no independent and identically distributed (i.i.d.) property is assumed for $N+1$ copies.

When a positive operator $\Omega$ satisfies the condition 
\begin{align}
\Omega \ge |G\rangle\langle G|,\label{TVG}
\end{align}
we define the spectral gap
$\nu(\Omega):= 1- \|\Omega - |G\rangle\langle G|\|$, 
where $\|A\|:=\lambda_{\rm max}(|A|)$,
$|A|:=\sqrt{A^{\dag}A}$, and $\lambda_{\rm max}(|A|)$ is the maximum eigenvalue of $|A|$.
Here, we consider the test operator $\Omega:=\sum_i\lambda_i\Pi_i$, where $\{\Pi_i\}_i$ are mutually orthogonal rank-one projectors with $\Pi_1=|G\rangle\langle G|$.
Since $\Omega(\le I)$ is a positive semidefinite operator and satisfies Eq.~\eqref{TVG}, $\lambda_1=1$ and $\{\lambda_i\}_{i\neq 1}$ are non-negative reals less than or equal to one.
Therefore, 
$\nu(\Omega)=\lambda_1-(\max_{i\neq 1}\lambda_i)$ is indeed the gap.
Hereafter, we only consider the case that $\nu(\Omega)>0$ holds.
Then, the paper \cite{ZH} showed the following.

\begin{proposition}[\protect{\cite[Theorem 1]{ZH}}]\label{PRE}
Assume that $\Omega$ satisfies Eq.~\eqref{TVG} and $\beta \ge \frac{1}{N\nu(\Omega)+1}$.
When the $N$-random sampling test of $\Omega$ 
 is passed,
the resultant state $\sigma$ satisfies 
 \begin{align}
\langle G|\sigma |G\rangle
\ge 1-  \frac{1-\beta}{N \beta \nu(\Omega)}
\label{TUGS}
\end{align}
with significance level $\beta$.
\end{proposition}

As the special case with $\nu(\Omega)=1$, we have the following proposition.
\begin{proposition}\label{CRST}
Assume that $\beta\ge \frac{1}{N+1} $. 
We consider $N+1$ binary variables $X_1, \ldots, X_{N+1}$.
We randomly choose $N$ variables from the above.
When all the $N$ values are zero,
the remaining variable $X'$ satisfies
\begin{align}
{\rm Pr} \{ X'=1\} \le \frac{1-\beta}{\beta N} 
\end{align}
with significance level $\beta$.
\end{proposition}

Notice that Proposition \ref{CRST} holds for any $N+1$ binary variables $X_1, \ldots, X_{N+1}$
whatever physical device generates the variables $X_1, \ldots, X_{N+1}$.
This is because Proposition \ref{CRST} is a statement with respect to 
the joint distribution among the variables $X_1, \ldots, X_{N+1}$.

\section{Adaptive Protocol with Perfect Match}\label{S2}
First, we assume that the verifier can choose the measurement basis dependently on the previous measurement outcomes.
Also, it is assumed that the verifier can choose any basis with the form
$\{|\alpha\rangle,|\alpha+\pi\rangle\}$, where
\begin{align}
|\alpha\rangle:=
\frac{1}{\sqrt{2}}(|0\rangle+e^{i\alpha}|1\rangle).
\end{align}

Based on an independence cover $\mathscr{A}=\{A_1, A_2, \ldots, A_m \}$ of $V$,
we construct the test operator $\Omega(\mathscr{A})$ satisfying Eq.~\eqref{TVG} as
\begin{align}
\Omega(\mathscr{A}):= \sum_{l=1}^m \frac{P_l}{m}.
\end{align}

The definition of the projection $P_l$ is given as follows.
First, 
the verifier measures any vertex $j \in A_l^c$ in the $Z$ basis and obtains the outcome $Z_j$.
Here, the superscript $c$ represents the complementary set.
By using the outcomes $\bm{Z}_l:= (Z_j)_{j \in A_l^c}$,
the expected state on the vertex $k \in A_l$ is given as $|\alpha_k(\bm{Z}_l)\rangle$,
where
\begin{align}
\alpha_k(\bm{Z}_l):= \sum_{j \in C_k} \theta_{j,k} Z_j
\end{align}
and $C_k$ is the set of vertices connected to the vertex $k$.
Then, 
the verifier measures any vertex $k \in A_l$ in the basis 
$\{|\alpha_k(\bm{Z}_l)\rangle,|\alpha_k(\bm{Z}_l)+\pi \rangle\}$.
When all the outcomes in $A_l$ correspond to 
$\otimes_{k \in A_l} |\alpha_k(\bm{Z}_l)\rangle$,
the verifier accepts the resultant state $\sigma$.
That is, using 
$Q_k:= \oplus_{\bm{z}_l} |\alpha_k(\bm{z}_l)\rangle_k~_k\langle\alpha_k(\bm{z}_l)|
\otimes |\bm{z}_l\rangle_{A_l^c}~_{A_l^c}\langle \bm{z}_l|$, we define
$P_l := \prod_{k \in A_l} Q_k $.

Hence, the operator $\Omega(\mathscr{A})$ satisfies Eq.~\eqref{TVG}.
For a subset $ B\subset [m]:=\{1, \ldots, m\}$, we
define the projection 
$P(B):=[\prod_{k \in B^c}(I-P_k)] (\prod_{j \in B}P_j) $.
Since $P([m])= |G\rangle \langle G|$, 
we have
\begin{align}
& \|\Omega(\mathscr{A})- |G\rangle \langle G|\|
=
\left\| \sum_{l=1}^m \frac{1}{m}( P_l- |G\rangle \langle G|)\right\|\nonumber \\
&=
\left\|\sum_{B \subsetneq [m]} \frac{|B|}{m} P(B)\right\|
=\frac{m-1}{m},
\end{align}
which implies that 
\begin{align}
\nu(\Omega(\mathscr{A}))=\frac{1}{m} .\label{NBT}
\end{align}
Here, $|B|$ represents the number of elements of $B$.
Hence, applying Proposition \ref{PRE},
we have the following theorem.

\begin{theorem}
\label{Theorem1}
The state $|G\rangle^{\otimes (N+1)}$
passes the $N$-random sampling test of
$\Omega(\mathscr{A})$
with probability $1$.
When the test is passed,
the resultant state $\sigma$ satisfies 
 \begin{align}
\langle G|\sigma |G\rangle
\ge 1-  \frac{m(1-\beta)}{N\beta}
\label{KVGS}
\end{align}
with significance level $\beta$.
\end{theorem}

\section{Adaptive Protocol with Imperfect Match}\label{S3}
Next, 
we assume that 
while the verifier can choose the measurement basis dependently on the previous measurement outcomes,
available bases for the verifier are limited to the following $h$ bases
$\{| \frac{\pi}{h}\rangle,| \frac{\pi}{h}+\pi \rangle\},
\{| \frac{2\pi }{h}\rangle,| \frac{2 \pi}{h}+\pi \rangle\},\ldots,
\{| \frac{h\pi }{h}\rangle,| \frac{h\pi }{h}+\pi \rangle\}$ for a positive integer $h$.

For an independence cover $\mathscr{A}=\{A_1, A_2, \ldots, A_m \}$ of $V$,
we define the test operator $\Omega_h(\mathscr{A})$ by modifying the test operator $\Omega(\mathscr{A})$ as
follows. 
First, we define $\alpha^h_k(\bm{Z}_l)$ as
$\frac{k \pi}{h}$ satisfying 
$ \frac{k \pi}{h}- \frac{\pi}{2h} \le \alpha_k(\bm{Z}_l) < \frac{k \pi}{h}+\frac{\pi}{2h}$.
Then, 
we define the operator $\Omega_h(\mathscr{A})$ 
and $P_{l;h}$
by 
replacing
the basis $\{|\alpha_k(\bm{Z}_l)\rangle,|\alpha_k(\bm{Z}_l)+\pi \rangle\}$ by 
the basis 
$\{|\alpha^h_k(\bm{Z}_l)\rangle,|\alpha^h_k(\bm{Z}_l)+\pi \rangle\}$
in the definitions of $\Omega(\mathscr{A})$ and $P_{l}$ in Sec.~\ref{S2}.

Unfortunately, 
the operator $\Omega_h(\mathscr{A})$ does not necessarily satisfy Eq.~\eqref{TVG}.
Instead, we have the following lemma.
\begin{lemma}
Let $|A_l|$ be the number of elements of $A_l$.
Then, we have the following evaluations.
 \begin{align}
&\langle G|\Omega_h(\mathscr{A}) |G\rangle
\ge 
\left( 1- \sin^2 \frac{\pi}{4h}\right)^{\max_l |A_l|}\label{TVG26} \\
& \|\Omega_h(\mathscr{A})
-\Omega(\mathscr{A})\|
\le
\left(\sum_{l=1}^m \frac{|A_l|}{m} \right)\sin \frac{\pi}{4h}
\label{TVG45}
\end{align}
\end{lemma}

\noindent{\it Proof:}
Since $|\langle \alpha^h_k(\bm{Z}_l)|\alpha_k(\bm{Z}_l)\rangle|^2 \ge 
1- \sin^2 \frac{\pi}{4h}$,
using $P_{\bm{Z}_l}(\bm{Z}_l):={\rm Tr}\langle G|\bm{Z}_l\rangle_{A_l^{c}}~_{A_l^{c}}\langle \bm{Z}_l|G\rangle$,
we have
 \begin{align}
&\langle G|\Omega_h(\mathscr{A}) |G\rangle
=
\sum_{l=1}^m \frac{1}{m}
\langle G| P_{l;h} |G\rangle \nonumber \\
=&
\sum_{l=1}^m \frac{1}{m}
\sum_{\bm{z}_l}
P_{\bm{Z}_l}(\bm{z}_l)
\prod_{k \in A_l} |\langle \alpha^h_k(\bm{z}_l)|\alpha_k(\bm{z}_l)\rangle|^2  \nonumber \\
\ge &
\sum_{l=1}^m \frac{1}{m}
\left( 1- \sin^2 \frac{\pi}{4h}\right)^{|A_l|}
\ge 
\left( 1- \sin^2 \frac{\pi}{4h}\right)^{\max_l |A_l|}.\label{TVG26T}
\end{align}
Also, 
since
\begin{align}
\| |\alpha_k(\bm{Z}_l)\rangle \langle \alpha_k(\bm{Z}_l)|
-
|\alpha_k^h(\bm{Z}_l)\rangle \langle \alpha_k^h(\bm{Z}_l)|
\|
\le \sin \frac{\pi}{4h},
\end{align}
we have
\begin{align}
&\|P_l-P_{l;h}\| \nonumber\\
\le &
\Big\| \oplus_{\bm{z}_l} |\bm{z}_l\rangle_{A_l^c}~_{A_l^c} \langle\bm{z}_l|
\otimes
 \nonumber\\
 &\big( \otimes_{k \in A_l} |\alpha_k(\bm{z}_l)\rangle \langle \alpha_k(\bm{z}_l)|
-
\otimes_{k \in A_l} |\alpha_k^h(\bm{z}_l)\rangle \langle \alpha_k^h(\bm{z}_l) | \big)
\Big\| \nonumber\\
= &\sup_{\bm{z}_l}
\Big\| \otimes_{k \in A_l} |\alpha_k(\bm{z}_l)\rangle \langle \alpha_k(\bm{z}_l)|
-
\otimes_{k \in A_l} |\alpha_k^h(\bm{z}_l)\rangle \langle \alpha_k^h(\bm{z}_l)|
\Big\| \nonumber\\
\le& \sup_{\bm{z}_l} \sum_{k \in A_l}
\big\|  |\alpha_k(\bm{z}_l)\rangle \langle \alpha_k(\bm{z}_l)|
- |\alpha_k^h(\bm{z}_l)\rangle \langle \alpha_k^h(\bm{z}_l)|\big\|\nonumber\\
\le & \sup_{\bm{z}_l} \sum_{k \in A_l}
\sin \frac{\pi}{4h}
=|A_l| \sin \frac{\pi }{4h}.
\end{align}
Hence,
\begin{align}
& \|\Omega_h(\mathscr{A})
-\Omega(\mathscr{A})\|
\le
\sum_{l=1}^m \frac{1}{m} \|P_l-P_{l;h}\| \nonumber \\
\le& \left(\sum_{l=1}^m \frac{|A_l|}{m} \right)\sin \frac{\pi}{4h}
.\label{TVG45T}
\end{align}
\hspace{\fill}$\blacksquare$

Using Proposition \ref{CRST}, and Eqs.~\eqref{NBT}, \eqref{TVG26}, and \eqref{TVG45}, we have the following theorem.
\begin{theorem}\label{Th2}
Assume that $\beta\ge \frac{1}{N+1} $. 
The state $|G\rangle^{\otimes (N+1)}$
passes the $N$-random sampling test of
$\Omega_h(\mathscr{A})$
with probability at least $(1-\sin^2 \frac{\pi}{4h})^{N\max_l |A_l|}$.
When the test is passed,
the resultant state $\sigma$ satisfies 
 \begin{align}
\langle G|\sigma |G\rangle
\ge 
1- \left[ \frac{m(1-\beta)}{\beta N}+n \sin \frac{\pi}{4h} \right]
\label{TVGS}
\end{align}
with significance level $\beta$.
\end{theorem}

Before giving the proof of Theorem~\ref{Th2}, we consider the asymptotic case to evaluate our adaptive protocol.
When
$\frac{N\max_l|A_l|}{h^2} \to 0$ ,
the passing probability with the correct state $|G\rangle$
converges to one as 
\begin{align}
&\left(1-\sin^2 \frac{\pi}{4h}\right)^{N\max_l |A_l|}
\ge 1- N\max_l |A_l| \sin^2 \frac{\pi}{4h}\nonumber \\
&\cong 
1- N\max_l |A_l| \frac{\pi^2}{16h^2}
\to 1, \label{FUY}
\end{align}
which implies that the verifier does not mistakenly reject the correct state $|G\rangle$.
For example,
when $m=n$, i.e., each color has only one vertex,
we have $|A_l|=1$.
In this case, when $N=an$ and $h=bn$ with positive constants $a$ and $b$,
Eq.~\eqref{FUY} holds, and 
\begin{align}
\frac{m(1-\beta)}{\beta N}+n \sin \frac{\pi}{4h}  
\to 
\frac{1-\beta}{a \beta}+\frac{\pi}{4b}.
\end{align}
That is, in the asymptotic regime, we can guarantee
 \begin{align}
\langle G|\sigma |G\rangle
\ge 1-  
\left(\frac{1-\beta}{a \beta}+\frac{\pi}{4b}\right)
\label{TVGR}
\end{align}
with significance level $\beta$.

To realize $\langle G|\sigma |G\rangle
\ge 1-  \epsilon$,
$a$ and $b$ need to satisfy $\frac{1-\beta}{a \beta}+\frac{\pi}{4b}=\epsilon$, i.e.,
\begin{align}
a =\frac{1-\beta}{ \beta} \left(\epsilon - \frac{\pi}{4b} \right)^{-1},
\end{align}
which requires the condition $\epsilon>\frac{\pi}{4b}$.

Now, we give the proof of Theorem~\ref{Th2} as follows.\\
\noindent{\it Proof:}
The first statement immediately follows from Eq.~\eqref{TVG26}.
Let $F$ be the fidelity between $\sigma$ and $|G\rangle \langle G|$.
Then, 
\begin{align}
& \Tr \sigma \Omega_h(\mathscr{A})
\le 
\Tr \sigma \Omega(\mathscr{A})
+\Tr \sigma
\left| \Omega_h(\mathscr{A})
-\Omega(\mathscr{A})\right| \nonumber \\
\stackrel{(a)}{\le} 
 &
\Tr \sigma 
\left[|G\rangle \langle G|+\left(1-\frac{1}{m}\right)(I-|G\rangle \langle G|)\right]\nonumber \\
&+\left(\sum_{l=1}^m \frac{|A_l|}{m} \right)\sin \frac{\pi}{4h}  \nonumber \\
= &
F+ (1-F)\left(1-\frac{1}{m}\right)+\left(\sum_{l=1}^m \frac{|A_l|}{m} \right)\sin \frac{\pi}{4h} \nonumber \\
= &
1- \frac{1-F}{m} +\left(\sum_{l=1}^m \frac{|A_l|}{m} \right)\sin \frac{\pi}{4h} ,
\end{align}
where $(a)$ follows from the combination of Eqs.~\eqref{NBT} and \eqref{TVG45}.

We virtually consider the case when we apply the two-valued POVM $\{\Omega_h(\mathscr{A}),I-\Omega_h(\mathscr{A})\}$
to all the $N+1$ systems.
Then, we define the variable $X_i$ as the outcome of the $i$-th system.
Here, 
the outcome $0$ corresponds to the POVM $\Omega_h(\mathscr{A}) $
and the outcome $1$ does to the POVM $I-\Omega_h(\mathscr{A}) $.
Now, we apply Proposition \ref{CRST} to the $N+1$ binary variables $X_1, \ldots, X_{N+1}$ defined here.
Under this application, we have 
${\rm Pr} \{ X'_{N+1}=1| X_1'= \ldots= X_{N}'=0\} =\Tr \sigma (I-\Omega_h(\mathscr{A}))$.
Hence, when the test is passed,
Proposition \ref{CRST} guarantees that
\begin{align}
\Tr \sigma \Omega_h(\mathscr{A})
\ge 1- \frac{1-\beta}{\beta N}
\end{align}
holds with significance level $\beta$.
Hence, solving the inequality
\begin{align}
1- \frac{1-F}{m} +\left(\sum_{l=1}^m \frac{|A_l|}{m} \right)\sin \frac{\pi}{4h} 
\ge 1- \frac{1-\beta}{\beta N},
\end{align}
we have
\begin{align}
1-F& \le m\left[ \frac{1-\beta}{\beta N}+\left(\sum_{l=1}^m \frac{|A_l|}{m} \right)\sin \frac{\pi}{4h} \right]
\nonumber \\
&=\frac{m(1-\beta)}{\beta N}+n \sin \frac{\pi}{4h}
\end{align}
with significance level $\beta$,
which is the desired statement.
\hspace{\fill}$\blacksquare$

\section{Non-adaptive Protocol with Perfect Match}\label{S4}
To consider a verification method without adaptive basis choice,
we consider another type of test.
Given integers $\mathbf{h}=\{h(k)\}_{k \in [n]}$
and
an independence cover $\mathscr{A}=\{A_1, A_2, \ldots, A_m \}$ of $V$,
we define the test operator
\begin{align}
\bar{P}_{l}:= \prod_{k \in A_l} \left(\frac{1}{h(k)}Q_k +\frac{h(k)-1}{h(k)}I\right).\label{CUT}
\end{align}
Then, we define the operator
\begin{align}
\bar{\Omega}(\mathscr{A})_{\mathbf{h}}:=\sum_{l=1}^m
\frac{1}{m}\bar{P}_{l}\label{CUT2},
\end{align}
which satisfies Eq.~\eqref{TVG}.
We have the following lemma.
\begin{lemma}\label{BCIL}
The spectral gap of $\bar{\Omega}(\mathscr{A})_{\mathbf{h}}$ is calculated as
\begin{align}
\nu(\bar{\Omega}(\mathscr{A})_{\mathbf{h}})
=\frac{1}{m\max_{k \in [n]}h(k)}.\label{BCI}
\end{align}
\end{lemma}

\noindent{\it Proof:}
For a subset $ B\subset A_l$, we
define the projection $Q(B):=[\prod_{k \in A_l\setminus B}(I-Q_k)] (\prod_{j \in B}Q_j) $.
Then,
\begin{align}
&\bar{P}_{l}
= \prod_{k \in A_l} \left[Q_k +\frac{h(k)-1}{h(k)}(I-Q_k)\right ]\nonumber \\
=& P_l+\sum_{B \subsetneq A_l} \prod_{k \in A_l\setminus B} \frac{h(k)-1}{h(k)}Q(B).
\end{align}
Hence,
\begin{align}
&\|\bar{\Omega}(\mathscr{A})_{\mathbf{h}}-|G\rangle \langle G| \|\nonumber \\
=&\max_{l}
\left(\frac{m-1}{m}+\frac{1}{m}\left\| \sum_{B \subsetneq A_l} \prod_{k \in A_l\setminus B} \frac{h(k)-1}{h(k)}Q(B)\right\|
\right)\nonumber \\
=&\max_{l}
\left(\frac{m-1}{m}+\frac{1}{m} \max_{k \in A_l} \frac{h(k)-1}{h(k)}\right).
\end{align}
Hence, using the relation 
$\max_{l}\max_{k \in A_l}h(k)
=\max_{k \in [n]}h(k)$, we obtain Eq.~\eqref{BCI}.
\hspace{\fill}$\blacksquare$

Therefore, combining Proposition \ref{PRE} and Lemma \ref{BCIL},
we have the following theorem.

\begin{theorem}\label{Th3}
The state $|G\rangle^{\otimes (N+1)}$
passes the $N$-random sampling test of
$\bar{\Omega}(\mathscr{A})_{\mathbf{h}}$
with probability $1$.
When this test is passed,
the resultant state $\sigma$ satisfies 
 \begin{align}
\langle G|\sigma |G\rangle
\ge 1-  \frac{m(1-\beta)\max_{k \in [n]}h(k) }{N\beta}
\label{KVGS2}
\end{align}
with significance level $\beta$.
\end{theorem}

Next, we discuss a test whose measurement basis cannot be chosen dependently 
on the obtained outcomes.
Also, we assume that possible values of $\alpha_k(\bm{z}_l)$ for $k \in A_l$
belongs to one of $e(k)$ bases
$\{| \alpha_{k,1}\rangle,| \alpha_{k,1}+\pi \rangle\},
\{| \alpha_{k,2}\rangle,| \alpha_{k,2}+\pi \rangle\},\ldots,
\{| \alpha_{k,e(k)}\rangle,| \alpha_{k,e(k)}+\pi \rangle\}$,
where $0 \le \alpha_{k,j}< \pi $ for $j=1, \ldots, e(k)$.
In this case, we consider the following protocol by modifying ${\Omega}(\mathscr{A})$.

When the verifier chooses $A_l$,
the verifier randomly chooses a measurement basis 
$\{| \alpha_{k,F_k}\rangle,| \alpha_{k,F_k}+\pi \rangle\}$
from $e(k)$ bases 
$\{| \alpha_{k,1}\rangle,| \alpha_{k,1}+\pi \rangle\},
\{| \alpha_{k,2}\rangle,| \alpha_{k,2}+\pi \rangle\},\ldots,
\{| \alpha_{k,e(k)}\rangle,| \alpha_{k,e(k)}+\pi \rangle\}$
with probability $1/e(k)$
and measures each of vertices in $A_l$ in this measurement basis
while the verifier measures the remaining vertices in the $Z$ bases.
Then, 
given $F_k$ and $\bm{Z}_l$, 
we define the subset ${A}_{l;F_k,\bm{Z}_l} \subset A_l$
as the set of vertices $k \in A_l$ satisfying the condition that
the chosen basis $\{| \alpha_{k,F_k}\rangle,| \alpha_{k,F_k}+\pi \rangle\}$ is correct.
The verifier considers that the test is passed 
when the measurement outcome at any vertex $k \in {A}_{l;F_k,\bm{Z}_l}$ corresponds to $|\alpha_k(\bm{Z}_l)\rangle$.
Note that when ${A}_{l;F_k,\bm{Z}_l}=\emptyset$, the test is always passed.

Since the verifier chooses the correct basis with probability $1/e(k)$ for any $k \in A_l$.
the above test is given as the operator $\bar{\Omega}(\mathscr{A})_{\mathbf{e}}$,
where $\mathbf{e}=\{e(k)\}_{k \in [n]}$.
That is, Theorem \ref{Th3} gives the performance of this test.

\section{Non-adaptive Protocol with Imperfect Match}
\label{VII}
Next, we consider the case when 
adaptive basis choice is not allowed and 
possible values of $\alpha_l(\bm{Z}_l)$
cannot be limited to a subset with reasonable elements.
Given an integer $h$
and
an independence cover $\mathscr{A}=\{A_1, A_2, \ldots, A_m \}$ of $V$,
by using the operators
\begin{align}
\bar{P}_{l;h}
&:= \prod_{k \in A_l} \left(\frac{1}{h}Q_{k;h} +\frac{h-1}{h}I\right),\nonumber \\
Q_{k;h}
&:= \oplus_{\bm{z}_l} |\alpha^h_k(\bm{z}_l)\rangle \langle\alpha^h_k(\bm{z}_l)|
\otimes |\bm{z}_l\rangle_{A_l^c}~_{A_l^c} \langle\bm{z}_l|,
\end{align}
we define the test operator
\begin{align}
\bar{\Omega}_h(\mathscr{A}):=\sum_{l=1}^m
\frac{1}{m}\bar{P}_{l;h}.
\end{align}
Then, we have the following lemma.
\begin{lemma}\label{L2S}
When $h(k)=h$ for any $k$, we denote $\mathbf{h}$ by $h$.
Then, we have
 \begin{align}
&\langle G|\bar{\Omega}_h(\mathscr{A}) |G\rangle
\ge 
\left( 1- \frac{1}{h}\sin^2 \frac{\pi}{4h}\right)^{\max_l |A_l|},\label{TVG2} \\
&\nu(\bar{\Omega}(\mathscr{A})_h)=\frac{1}{mh},
\label{BFU}\\
&\|\bar{\Omega}_h(\mathscr{A})
-\bar{\Omega}(\mathscr{A})_h\|
\le \left(\sum_{l=1}^m \frac{|A_l|}{mh } \right)\sin \frac{\pi}{4h}
.\label{TVG5}
\end{align}
\end{lemma}

\noindent{\it Proof:}
Eq.~\eqref{TVG2} can be shown as follows.
 \begin{align}
&\langle G|\bar{\Omega}_h(\mathscr{A}) |G\rangle
=
\sum_{l=1}^m \frac{1}{m}
\langle G| \bar{P}_{l;h} |G\rangle \nonumber \\
=&
\sum_{l=1}^m \frac{1}{m}
\sum_{\bm{z}_l}
P_{\bm{Z}_l}(\bm{z}_l) \nonumber \\
&\prod_{k \in A_l} 
\langle \alpha_k(\bm{z}_l)|
\left(\frac{1}{h}|\alpha_k^h(\bm{z}_l)\rangle\langle \alpha_k^h(\bm{z}_l)| +\frac{h-1}{h}I\right)|\alpha_k(\bm{z}_l)\rangle  \nonumber \\
=&
\sum_{l=1}^m \frac{1}{m}
\sum_{\bm{z}_l}
P_{\bm{Z}_l}(\bm{z}_l)
\prod_{k \in A_l} 
\left(\frac{h-1}{h}+\frac{1}{h}
|\langle \alpha_k(\bm{z}_l)|\alpha_k^h(\bm{z}_l)\rangle|^2 \right) \nonumber \\
\ge &
\sum_{l=1}^m \frac{1}{m}
\left( 1- \frac{1}{h} \sin^2 \frac{\pi}{4h}\right)^{|A_l|}
\ge 
\left( 1- \frac{1}{h}\sin^2 \frac{\pi}{4h}\right)^{\max_l |A_l|}.\label{TVG2T}
\end{align}

Since $h(k)=h$, Lemma \ref{BCIL} implies
Eq.~\eqref{BFU}
and
\begin{align}
&\|\bar{P}_l-\bar{P}_{l;h}\| \nonumber\\
= &
\bigg\| \oplus_{\bm{z}_l} |\bm{z}_l\rangle_{A_l^c}~_{A_l^c} \langle\bm{z}_l|\otimes
 \nonumber\\
 &\bigg [ \otimes_{k \in A_l} 
 \left(\frac{1}{h} |\alpha_k(\bm{z}_l)\rangle \langle \alpha_k(\bm{z}_l)|
 + \frac{h-1}{h}I\right)
 \nonumber\\
&-
\otimes_{k \in A_l} \left(\frac{1}{h}|\alpha_k^h(\bm{z}_l)\rangle \langle \alpha_k^h(\bm{z}_l) |
 + \frac{h-1}{h}I\right)
\bigg ]
\bigg\| \nonumber\\
= &\sup_{\bm{z}_l}\bigg\| \otimes_{k \in A_l} 
 \left(\frac{1}{h} |\alpha_k(\bm{z}_l)\rangle \langle \alpha_k(\bm{z}_l)|
 + \frac{h-1}{h}I\right)
 \nonumber\\
&-
\otimes_{k \in A_l} \left(\frac{1}{h}|\alpha_k^h(\bm{z}_l)\rangle \langle \alpha_k^h(\bm{z}_l) |
 + \frac{h-1}{h}I\right)
\bigg\| \nonumber\\
\le & \sup_{\bm{z}_l} \sum_{k \in A_l}
\bigg\|
\left(\frac{1}{h} |\alpha_k(\bm{z}_l)\rangle \langle \alpha_k(\bm{z}_l)|
 + \frac{h-1}{h}I\right)
 \nonumber\\
&-
 \left(\frac{1}{h}|\alpha_k^h(\bm{z}_l)\rangle \langle \alpha_k^h(\bm{z}_l) |
 + \frac{h-1}{h}I\right)
\bigg\| \nonumber\\
= & \sup_{\bm{z}_l} \sum_{k \in A_l}\frac{1}{h}
\big\|  |\alpha_k(\bm{z}_l)\rangle \langle \alpha_k(\bm{z}_l)|
- |\alpha_k^h(\bm{z}_l)\rangle \langle \alpha_k^h(\bm{z}_l)|\big\|\nonumber\\
\le & \sup_{\bm{z}_l} \sum_{k \in A_l}\frac{1}{h}
\sin \frac{\pi}{4h}
=\frac{|A_l| }{h}\sin \frac{\pi }{4h}.
\end{align}
Hence,
\begin{align}
& \|\bar{\Omega}_h(\mathscr{A})
-\bar{\Omega}(\mathscr{A})_h\|
\le
\sum_{l=1}^m \frac{1}{m} \| \bar{P}_l-\bar{P}_{l;h}\| \nonumber \\
\le& \left(\sum_{l=1}^m \frac{|A_l|}{mh } \right)\sin \frac{\pi}{4h}
.\label{TVG5L}
\end{align}
\hspace{\fill}$\blacksquare$

Using Eqs.~\eqref{TVG2}, \eqref{BFU}, and \eqref{TVG5} of Lemma \ref{L2S},
we can show the following theorem in the same way as Theorem \ref{Th2}.
That is, it can be shown by replacing 
\eqref{TVG26}, \eqref{NBT}, and \eqref{TVG45}
in the proof of Theorem \ref{Th2}
by \eqref{TVG2}, \eqref{BFU}, and \eqref{TVG5}, respectively.

\begin{theorem}\label{Theorem4}
Assume that $\beta\ge \frac{1}{N+1} $. 
The state $|G\rangle^{\otimes (N+1)}$
passes the $N$-random sampling test of
$\bar{\Omega}_h(\mathscr{A})$ with probability at least $(1-\frac{1}{h}\sin^2 \frac{\pi}{4h})^{N\max_l |A_l|}$.
When the test is passed,
the resultant state $\sigma$ satisfies 
 \begin{align}
\langle G|\sigma |G\rangle
\ge 
1- \left[ \frac{mh(1-\beta)}{\beta N}+n \sin \frac{\pi}{4h} \right]
\label{TVGS2}
\end{align}
with significance level $\beta$.
\end{theorem}

Now, we construct a protocol to realize the test operator 
$\bar{\Omega}_h(\mathscr{A})$
without adaptive basis choice 
when 
possible values of $\alpha_l(\bm{Z}_l)$
cannot be limited to a subset with reasonable elements
The verifier randomly choose $A_l$ from 
an independence cover $\mathscr{A}=\{A_1, A_2, \ldots, A_m \}$ of $V$.
When the verifier chooses $A_l$,
the verifier randomly chooses the measurement basis 
$\{| \frac{F\pi}{h}\rangle,| \frac{F\pi}{h}+\pi \rangle\}$
from $h$ bases 
$\{| \frac{\pi}{h}\rangle,| \frac{\pi}{h}+\pi \rangle\},
\{| \frac{2\pi }{h}\rangle,| \frac{2 \pi}{h}+\pi \rangle\},\ldots,
\{| \frac{h\pi }{h}\rangle,| \frac{h\pi }{h}+\pi \rangle\}$ 
with probability $1/h$ and measures each of vertices in $A_l$ in this measurement basis
while the verifier measures the remaining vertices in the $Z$ bases.
Then, 
given $F$ and $\bm{Z}_l$, 
we define the subset ${A}_{l;h,F,\bm{Z}_l} \subset A_l$
as the set of vertices $k \in A_l$ satisfying the condition that
the chosen basis state 
$| \frac{F\pi }{h}\rangle$ or $| \frac{F\pi }{h}+\pi \rangle$ 
equals to the correct basis $|\alpha^h_k(\bm{Z}_l)\rangle$.
The verifier considers that the test is passed 
when the measurement outcome at any vertex $k \in {A}_{l;h,F,\bm{Z}_l}$ corresponds to $|\alpha^h_k(\bm{Z}_l)\rangle$.
Since the verifier chooses the correct basis with probability $1/h$ for any $k \in A_l$,
this test is given as the test operator 
$\bar{\Omega}_h(\mathscr{A})$.

For example,
when $m=n$, i.e., each color has only one vertex,
we have $|A_l|=1 $.
In this case, when $N=an^2$ and $h=bn$ with positive constants $a$ and $b$,
the passing probability with the correct state $|G\rangle$
is
\begin{align}
&\left(1-\frac{1}{h}\sin^2 \frac{\pi}{4h}\right)^{N}
=\left(1-\frac{1}{bn}\sin^2 \frac{\pi}{4bn}\right)^{an^2}
\nonumber \\
\ge & 1-\frac{an^2}{bn}\sin^2 \frac{\pi}{4bn}
\cong 1- \frac{a \pi^2}{16b^3n}
= 1- o\left(\frac{1}{n}\right).
\end{align}
On the other hand,
\begin{align}
\frac{mh(1-\beta)}{\beta N}+n \sin \frac{\pi}{4h}  
\to 
\frac{(1-\beta)b}{a \beta}+\frac{\pi}{4b}.
\end{align}
That is, in the asymptotic regime, we can guarantee
 \begin{align}
\langle G|\sigma |G\rangle
\ge 1-  
\left[\frac{(1-\beta)b}{a \beta}+\frac{\pi}{4b}\right]
\label{TVGR2}
\end{align}
with significance level $\beta$.

To realize $\langle G|\sigma |G\rangle
\ge 1-  \epsilon$,
$a$ and $b$ need to satisfy $\frac{(1-\beta)b}{a \beta}+\frac{\pi}{4b}=\epsilon$, i.e.,
\begin{align}
a =\frac{1-\beta}{ \beta} \left(\frac{\epsilon}{b} - \frac{\pi}{4b^2} \right)^{-1}.
\end{align}
The function $b \mapsto (\frac{\epsilon}{ b} - \frac{\pi}{4b^2} )^{-1}$
realizes the minimum value $\frac{\pi}{\epsilon^2} $
when $b= \frac{\pi}{2\epsilon}$.
That is, when $\epsilon=\frac{\pi}{2b}$, $N=\frac{ \pi(1-\beta)}{\beta\epsilon^2} n^2 $ is sufficient to guarantee
$\langle G|\sigma |G\rangle
\ge 1-  \epsilon$
with significance level $\beta$
in the asymptotic regime.

\section{Applications}
\label{VIII}
In this section, we apply our verification protocols to verify several quantum computing models. In Sec.~\ref{VIIIA}, we consider the verification of the MBQC~\cite{RB01,RBB03}. In Sec.~\ref{VIIIB}, we consider the verification of IQP circuits~\cite{SB09}. Although all of our verification protocols can be applied to these purposes, for simplicity, we focus on our third protocol proposed in Sec.~\ref{S4}.
\subsection{Verification of measurement-based quantum computing}
\label{VIIIA}
MBQC~\cite{RB01,RBB03} is one of the most promising universal quantum computing models.
In MBQC, quantum computing proceeds by adaptively measuring each qubits of an entangled state, a so-called universal resource state.
So far, several universal resource states have been proposed~\cite{BR01,RHG06,KW17,TMH18}.
Among them, the M{\o}lmer-S{\o}rensen (MS) graph state~\cite{KW17}
\begin{eqnarray}
\label{MS}
|G_{\rm MS}\rangle:=\left(\prod_{(i,j)\in E}e^{-i\theta_{ij}Z_i\otimes Z_j}\right)|+\rangle^{\otimes n}
\end{eqnarray}
with $\theta_{ij}\in\{\frac{\pi}{8},\frac{\pi}{4}\}$
is particularly attractive. This is because only $X$ and $Z$-basis measurements are sufficient to perform MBQC on the MS graph state. From Eq.~\eqref{MS}, MS graph states are weighted graph states up to local (single-qubit) unitary transformations $\prod_{i\in V}U_i$. Therefore, by transforming the measurement basis on the $i$-th vertex by $U_i^{\dag}$ in our verification protocol, we can apply our protocol to estimate the fidelity between the MS graph state and a quantum state generated by experiment.
In the case of the MS graph state,
$\max_{k \in [n]}e(k)\le 8$ and $m\le n$.
Hence,
\begin{eqnarray}
N=\cfrac{8n(1-\beta)}{\epsilon\beta}
\end{eqnarray}
is sufficient to guarantee $\langle G_{\rm MS}|\sigma|G_{\rm MS}\rangle\ge 1-\epsilon$ with significance level $\beta$.

\begin{figure}[t]
\includegraphics[width=8cm, clip]{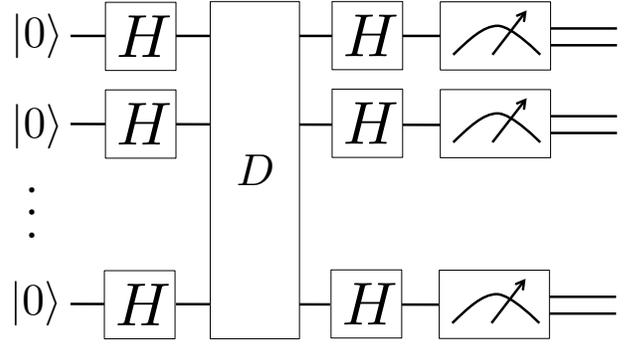}
\caption{An IQP circuit. $H$ and $D$ represent the Hadamard gate and a $Z$-diagonal gate, respectively. Meter symbols represent the $Z$-basis measurements.}
\label{IQPfig}
\end{figure}

\subsection{Verification of instantaneous quantum polynomial time circuits}
\label{VIIIB}
\begin{figure}[t]
\includegraphics[width=8cm, clip]{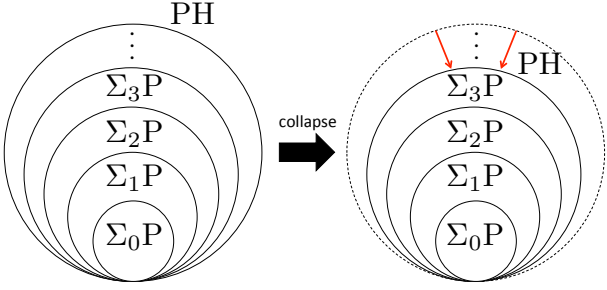}
\caption{Illustration of the collapse of PH to its third level $\Sigma_3 {\rm P}$. The ${\rm PH}=\cup_{i\ge 0}\Sigma_i{\rm P}$ is an infinite tower of complexity classes, where $\Sigma_i{\rm P}$ represents the $i$-th level of the PH. If all levels above the third is contained in the third level, we say that the PH collapses to its third level.}
\label{PH}
\end{figure}

In this subsection, we consider the verification of quantum supremacy demonstrations with IQP circuits~\cite{SB09}.
An $n$-qubit IQP circuit is defined as follows (see Fig.~\ref{IQPfig}).
\begin{definition}[IQP]
\label{IQP}
An $n$-qubit IQP circuit is a quantum circuit that satisfies following conditions
\begin{enumerate}
\item The initial state is $|0\rangle^{\otimes n}$.
\item The $n$-qubit unitary $H^{\otimes n}DH^{\otimes n}$ is applied, where $H$ is the Hadamard gate, and $D$ is a unitary consisting of polynomial number of $Z$-diagonal gates.
\item Finally, all of $n$ qubits are measured in the $Z$ bases.
\end{enumerate}
\end{definition}
From Definition~\ref{IQP}, the IQP circuit does not seem to be a universal quantum computing model.
However, the hardness of classically simulating the IQP circuits has been shown under a certain unproven conjecture. To explain this fact in more detail, we use the following definition.
\begin{definition}
Let $\{q_z\}_z$ be the output probability distribution of an $n$-qubit quantum circuit $Q_n$. If there exists a ${\rm poly}(n)$-time classical sampler whose output probability distribution $\{p_z\}_z$ satisfies
\begin{eqnarray}
\sum_z|q_z-p_z|\le \delta,
\end{eqnarray}
we say that the output probability distribution $\{q_z\}_z$ of $Q_n$ is classically simulated in ${\rm poly}(n)$ time with an $l_1$-norm error $\delta$.
\end{definition}
\noindent Bremner, Montanaro, and Shepherd have shown that, assuming a certain unproven conjecture, output probability distributions of IQP circuits cannot be classically simulated in ${\rm poly}(n)$ time with a constant $l_1$-norm error unless the polynomial-time hierarchy (PH) collapses to its third level~\cite{BMS16}.
The PH is an infinite tower of complexity classes. In other words, when we write the $i$-th level of the PH as a complexity class $\Sigma_i{\rm P}$, PH$=\cup_{i\ge 0}\Sigma_i{\rm P}$ (for more formal definition, see Ref.~\cite{P94}). If PH$\subseteq\Sigma_i{\rm P}$, we say that the PH collapses to its $i$-th level (see Fig.~\ref{PH}). In the field of computer science, it is widely believed that the PH does not collapse.
Therefore, their result suggests the quantum computational advantage of IQP circuits, a so-called quantum (computational) supremacy.

More precisely, they have shown the following theorem.
\begin{theorem}[\cite{BMS16}]
\label{supremacy}
Assume either one of below two conjectures is true. If the output probability distribution of any IQP circuit can be classically simulated in polynomial time, up to an error of $\frac{1}{192}$ in $l_1$ norm, then the PH would collapse to its third level.
\end{theorem}
\begin{conjecture}[\cite{BMS16}]
\label{partition}
Let
\begin{eqnarray}
Z_R:=\sum_{z\in\{\pm 1\}^n}e^{i\pi/8\left(\sum_{j<k}w_{jk}z_jz_k+\sum_{l=1}^nv_lz_l\right)},
\end{eqnarray}
where $j,k\in\{1,2,\ldots,n\}$ and $w_{jk}, v_{l}\in\{0,1,\ldots,7\}$.
It is ${\rm \# P}$-hard to approximate $|Z_R|^2$ up to a multiplicative error $\frac{1}{4}+o(1)$ for a $\frac{1}{24}$ fraction of instances over the choice of $\{w_{jk}\}_{j<k}$ and $\{v_l\}_{l=1}^n$.
\end{conjecture}
\begin{conjecture}[\cite{BMS16}]
\label{gap}
Let $f:\{0,1\}^n\rightarrow\{0,1\}$ be a uniformly random degree-three polynomial over $\mathbb{F}_2$. Then, it is $\# {\rm P}$-hard to approximate $(\frac{{\rm gap}(f)}{2^n})^2$ up to a multiplicative error of $\frac{1}{4}+o(1)$ for a $\frac{1}{24}$ fraction of polynomials $f$. Here, ${\rm gap}(f):=|\{x:f(x)=0\}|-|\{x:f(x)=1\}|$.
\end{conjecture}
Here, we say that a function $g$ is approximated up to multiplicative error $\delta$ if $g'$ is obtained such that $|g-g'|\le\delta g$ holds.
$\#{\rm P}$~\cite{V79} is a class of function problems that can be solved by counting the number of solutions of arbitrary ${\rm NP}$ problems.

When we assume that Conjecture~\ref{gap} is true, Theorem~\ref{supremacy} holds for the IQP circuits whose diagonal gate $D$ is composed of $Z$, the controlled-$Z$, and the controlled-controlled-$Z$ gates. In this case, output states of IQP circuits (immediately before the $Z$-basis measurements) are hypergraph states~\cite{RHBM13}, which are generalizations of graph states, up to local unitary transformations. Therefore, such the IQP circuits can be verified using existing polynomial-time verification protocols for hypergraph states~\cite{TM18,ZH}.

However, since Conjecture~\ref{gap} has not yet been shown, there is a possibility that Conjecture~\ref{gap} is incorrect. That is why it is important to consider the case that Conjecture~\ref{partition} is true. When Conjecture~\ref{partition} is true, Theorem~\ref{supremacy} holds for the IQP circuits whose diagonal gate $D$ is composed of $T:=|0\rangle\langle0|+e^{i\pi/4}|1\rangle\langle 1|$ and $\Lambda(\frac{\pi}{2})$. Therefore, the output state of the IQP circuit is
\begin{eqnarray}
\nonumber
&&|G_{\rm IQP}\rangle\\
\nonumber
&:=&\left(\prod_{l=1}^nH_lT_l^{v_l}\right)\left[\prod_{j<k}{T_j^\dag}^{w_{jk}}{T_k^{\dag}}^{w_{jk}}\Lambda_{jk}\left(\frac{w_{jk}\pi}{2}\right)\right]|+\rangle^{\otimes n}\\
{}
\end{eqnarray}
that is a weighted graph state up to local unitary transformations.
Since the IQP model can be considered as MBQC with non-adaptive measurements, we should not use adaptive measurements to verify the output state $|G_{\rm IQP}\rangle$. Therefore, we focus on our third verification protocol in this subsection.
Since ${\rm max}_{k\in[n]}e(k)\le 2$ and $m\le n$, by using our third protocol,
\begin{eqnarray}
N=\cfrac{2n(1-\beta)}{\epsilon\beta}
\end{eqnarray}
is sufficient to guarantee $\langle G_{\rm IQP}|\sigma|G_{\rm IQP}\rangle\ge 1-\epsilon$ with significance level $\beta$.

At the last of this subsection, we show that a quantum state $\sigma$ that passes our verification protocol can be used to demonstrate the quantum supremacy.
To this end, from Theorem~\ref{supremacy}, we show the following corollary.
\begin{corollary}
\label{supremacy2}
Assume Conjecture~\ref{partition} is true. 
If 
for any output state $|G_{\rm IQP}\rangle$, 
there exists an $n$-qubit quantum state $\sigma$ such that 
$\langle G_{\rm IQP}|\sigma|G_{\rm IQP}\rangle\ge 1-\epsilon$ with $\epsilon=\frac{1}{{\rm poly}(n)}$,
and
the probability 
distribution $\{\langle z|\sigma|z\rangle\}_{z\in\{0,1\}^n}$ can be classically simulated in polynomial time, up to an error of $\frac{1}{193}$ in $l_1$ norm, 
then the PH would collapse to its third level.
\end{corollary}
{\it Proof:}
Let $F$ be the fidelity between $\sigma$ and $|G_{\rm IQP}\rangle$. Then, we have
\begin{eqnarray}
\nonumber
&&\sum_{z\in\{0,1\}^n}\left||\langle z|G_{\rm IQP}\rangle|^2-\langle z| \sigma|z\rangle\right|\le 2\sqrt{1-F}\\
\label{51}
&&\le 2\sqrt{\epsilon}=\cfrac{1}{{\rm poly}(n)}.
\end{eqnarray}
Let $p_z$ be the probability of a classical sampler outputting $z$.
Then, if we assume that it is possible to classically simulate the probability distribution $\{\langle z|\sigma|z\rangle\}_{z\in\{0,1\}^n}$ in polynomial time, up to an error of $\frac{1}{193}$ in $l_1$ norm, from the triangle inequality and Eq.~\eqref{51},
\begin{eqnarray}
\nonumber
&&\sum_{z\in\{0,1\}^n}\left||\langle z|G_{\rm IQP}\rangle|^2-p_z\right|\\
\nonumber
&\le&\sum_{z\in\{0,1\}^n}\left||\langle z|G_{\rm IQP}\rangle|^2-\langle z| \sigma|z\rangle\right|+\sum_{z\in\{0,1\}^n}\left|\langle z| \sigma|z\rangle-p_z\right|\\
&\le&\cfrac{1}{{\rm poly}(n)}+\cfrac{1}{193}\le\cfrac{1}{192}.
\end{eqnarray}
This consequence means that it is possible to classically simulate the output probability distribution of the IQP circuit in polynomial time, up to an error of $\frac{1}{192}$ in $l_1$ norm. Therefore, from Theorem~\ref{supremacy}, the PH collapses to its third level.
\hspace{\fill}$\blacksquare$

From Theorem~\ref{Th3}, using $N=\frac{2n(1-\beta)}{\epsilon\beta}$ copies, 
with significance level $\beta$,
we can prepare an $n$-qubit quantum state $\sigma$ whose fidelity with $|G_{\rm IQP}\rangle$ is at least $1-\epsilon$.
When $\epsilon,\beta=\frac{1}{{\rm poly}(n)}$, $N={\rm poly}(n)$, i.e., this preparation can be accomplished in polynomial time. Therefore, by measuring the quantum state $\sigma$ in the $Z$ basis, it is possible to generate the probability distribution $\{\langle z|\sigma|z\rangle\}_{z\in\{0,1\}^n}$ in polynomial time. On the other hand, from Corollary~\ref{supremacy2}, when we assume that the PH does not collapse, this is impossible for any classical sampler. This means that the quantum state $\sigma$ that passes our (third) verification protocol can be used to demonstrate the quantum supremacy.

\section{Conclusion \& discussion}
\label{IX}
We have proposed four kinds of verification protocols of weighted graph states for each of the following classes of measurements: (i) adaptive and all bases are available, (ii) adaptive and restricted bases are available, (iii) non-adaptive and all bases are available, (iv) non-adaptive and restricted bases are available. The comparison of Theorems~\ref{Theorem1}, \ref{Th2}, \ref{Th3}, and \ref{Theorem4} yields the relationships among these four protocols. As far as we know, so far, no efficient verification protocol has been proposed for weighted graph states. Applying our protocols, we have also shown that the MBQC and the IQP model can be efficiently verified.

In our verification protocols, we assume that the verifier's single-qubit measurements are ideal.  One possible solution to remove this assumption is to utilize the quantum error correction. In Ref.~\cite{FH17}, the Raussendorf-Harrington-Goyal (RHG) lattice state~\cite{RHG07} enables the verifier to do the topological quantum error correction with only physical single-qubit measurements during the verification of the universal MBQC. Unfortunately, such a scheme is known only for graph states. If a similar scheme is found for weighted graph states, we may be able to add the fault tolerance in our verification protocols.

As another possible solution to remove the assumption, we can consider a classical verification protocol that requires no quantum operation for the verifier. In Ref.~\cite{HKEG18}, under some assumptions, Hangleiter {\it et al.} have shown that this approach requires exponentially many runs of the IQP circuit. To circumvent this no-go result, the self-testing approach may be helpful. So far, several self-testing protocols have been proposed for maximally entangled pair of qubits~\cite{MY04, HH18}, graph states~\cite{HH18, M14}, the three-qubit W state~\cite{WCYLBS14}, and all pure bipartite entangled states~\cite{CGS17}. It is an interesting future work to propose a self-testing protocol for weighted graph states.

\section*{ACKNOWLEDGMENTS}
We thank Tomoyuki Morimae and Yasuhiro Takahashi for helpful discussions.
M. H. is supported in part by Fund for the Promotion of Joint International Research (Fostering Joint International Research) Grant No. 15KK0007, Japan Society for the Promotion of Science (JSPS) Grant-in-Aid for Scientific Research (A) No. 17H01280, (B) No. 16KT0017, and Kayamori Foundation of Informational Science Advancement.
Y. T. is supported by MEXT QLEAP project.

\end{document}